\documentclass[usegraphicx]{mn2e}
\usepackage{times}

\title[Discovery of a remarkable subpulse drift pattern in PSR B0818$-$41]
{Discovery of a remarkable subpulse drift pattern in PSR B0818$-$41}
\author[B. Bhattacharyya]
{B. Bhattacharyya,$^1$
 Y. Gupta$^1$, J. Gil$^2$, M. Sendyk$^2$\\\\ 
 $^1$National Centre for Radio Astrophysics, TIFR, Pune University Campus, Post Bag 3, 
Pune 411 007, India\\
 $^2$Institute of Astronomy, University of Zielona Gora, Lubuska 2, 65-265 Zielona Gora, Poland }
\date{Accepted. Received}
\begin{document}
\label{firstpage}
\maketitle

\pagerange{\pageref{firstpage}--\pageref{lastpage}} \pubyear{2006}

\def\LaTeX{L\kern-.36em\raise.3ex\hbox{a}\kern-.15em
    T\kern-.1667em\lower.7ex\hbox{E}\kern-.125emX}
\begin{abstract}
We report the discovery of a remarkable subpulse drift pattern in the relatively 
less studied wide profile pulsar, B0818$-$41, using high sensitivity GMRT observations. 
We find simultaneous occurrence of three drift regions with two different drift rates: 
an inner region with steeper apparent drift rate flanked on each side by a region of 
slower apparent drift rate. Furthermore, these closely spaced drift bands always maintain 
a constant phase relationship. Though these drift regions have significantly different 
values for the measured $P_2$, the measured $P_3$ value is the same and equal to $18.3\:{P_1}$.  
We interpret the unique drift pattern of this pulsar as being created by the intersection 
of our line of sight (LOS) with two conal rings on the polar cap of a fairly aligned 
rotator (inclination angle $\alpha \sim 11$\degr), with an ``inner'' LOS geometry (impact 
angle $\beta \sim -5.4$\degr). We argue that both the rings have the same values for the 
carousel rotation periodicity $P_4$ and the number of sparks $N_{sp}$. We find that $N_{sp}$ 
is 19-21 and show that it is very likely that, $P_4$ is the same as the measured $P_3$, making 
it a truly unique pulsar. We present results from simulations of the radiation pattern using
the inferred parameters, that support our interpretations and reproduce the average profile 
as well as the observed features in the drift pattern quite well. 
\end{abstract}

\section{Introduction}                \label{sec:intro}    
The study of pulsars showing systematic subpulse drift patterns provides important clues 
for the understanding of the unsolved problem of pulsar emission mechanism. Constraints 
provided by such observations can have far reaching implications for the theoretical models, 
as exemplified by some of the recent results in this area (e.g. \cite {Desh_etal} and \cite 
{Gupta_etal}). In this context, pulsars with wide profiles $-$ interpreted as, emission 
coming from one magnetic pole of a highly aligned pulsar (pulsar with the magnetic axis 
almost parallel to the rotation axis) $-$ can provide extra insights because of the 
presence of multiple drift bands, as illustrated in the recent studies of B0826$-$34 (\cite 
{Gupta_etal} and \cite {Esamdin_etal}) and B0815$+$09 (\cite {Qiao_etal_04}).

B0818$-$41 is a relatively less studied wide profile pulsar with emission occurring for more 
than 180\degr of pulse longitude. Discovered during the second Molonglo pulsar survey 
(\cite {Manchester_etal}), it has a period of 0.545 $s$ and is relatively old, with a 
characteristic age of $4.57\times10^{8}$ years. The inferred dipolar magnetic field of this 
pulsar is $1.03\times10^{11}$ G, which is a typical value for slow pulsars.  From a study of 
its average polarization behaviour at 660 and 1440 MHz, \cite {Qiao_etal} predict that the 
pulsar must have a small inclination angle between the magnetic and rotation axes. 

We have carried out high sensitivity GMRT observations of B0818$-$41, which bring out a unique 
pattern of subpulse drift, which is hitherto not reported. The main results from our initial 
observations and the interpretations thereof are reported in this paper. 
     
\section{Observations and data analysis}                   \label{sec:observations} 
We observed B0818$-$41 at 325 MHz, on 24th February, 2003 at the GMRT, using the phased 
array mode (\cite {Gupta_etal_00}). The raw data were recorded with 0.512 msec sampling 
interval. During the offline analysis we further integrated the raw data to achieve 
the final time resolution of 2.048 msec. The duration of the observations was about 
31 minutes (i.e. 3414 pulses).
\begin{figure}  
\begin{center}
\includegraphics[angle=-90,width=0.33\textwidth]{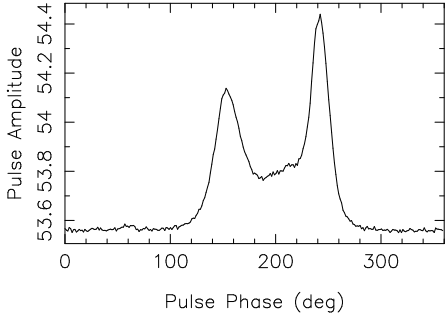}
\caption{Average profile of PSR B0818$-$41 at 325 MHz. The pulse amplitude is in 
arbitrary units.}
\label{fig:avg_prof}
\end{center}
\end{figure} 
\begin{figure}      
\begin{center}
\includegraphics[angle=0, width=0.4\textwidth]{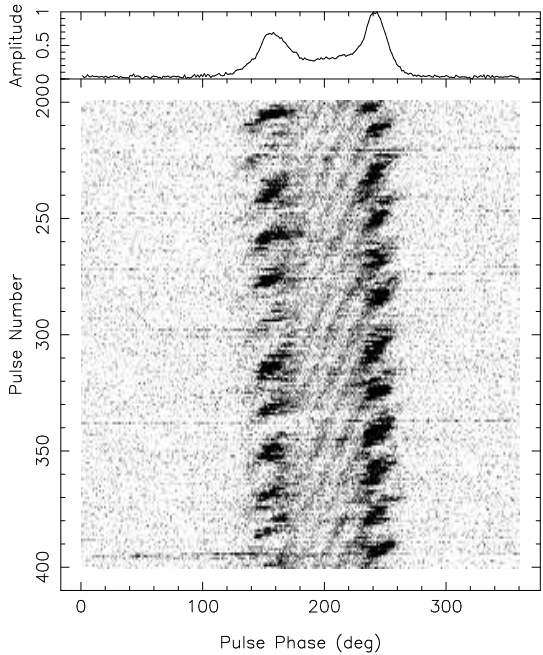}
\caption{Gray scale plot of single pulse data (pulse \# 200 to 400) of PSR B0818$-$41 
at 325 MHz, with the average profile shown on top. Signatures of radio frequency 
interference are present around pulse number 220, 298, 338 and 397.}
\label{fig:sp_200_400}
\end{center}
\end{figure}    
\begin{figure}    
\begin{center}
\includegraphics[angle=0, width=0.4\textwidth]{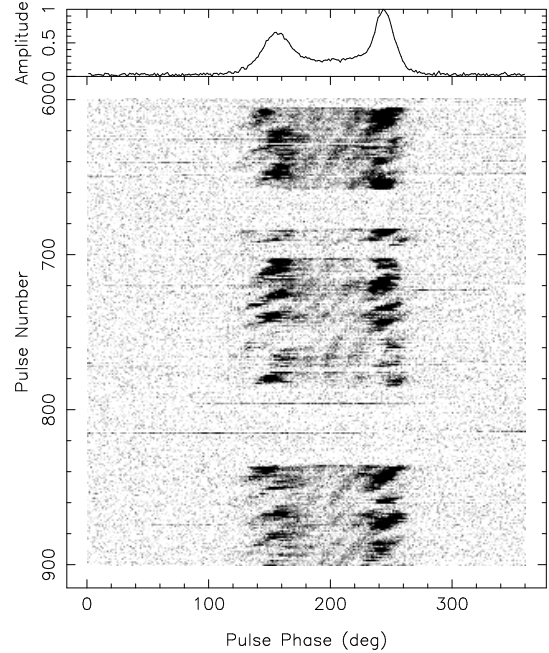}
\caption{Gray scale plot of single pulse data (pulse \# 600 to 900) of PSR B0818$-$41 
at 325 MHz, with the average profile shown on the top. Signatures 
of radio frequency interference are present around pulse number 625, 720, 790 and 815.}
\label{fig:sp_600_900}
\end{center}
\end{figure}         

The raw data were first dedispersed (with a DM value equal to 113.4 $pc/cm^3$) and then bad 
data points were filtered out from the dedispersed data. The final average pulse profile, 
illustrated in Fig. \ref{fig:avg_prof}, was obtained by synchronous folding of this data 
with the Doppler corrected pulsar period. It shows a fairly steep sided and double peaked 
structure. The two components of the double peaked profile are separated by 
86.6\degr$\pm$0.3\degr and are joined by a saddle region.

Single pulse sequences from our data reveal an interesting pattern of drifting. 
Fig. \ref{fig:sp_200_400} shows a typical drift pattern over a sequence of 200 pulses. We 
see three different drift regions with two different drift rates $-$ an inner region with 
steeper apparent drift rate flanked on each side by a region of slower apparent drift rate. 
Multiple drift bands (typically 3 to 4) can be seen in the inner region for any single pulse, 
whereas the outer regions show only a single drift band. Though the regular drift patterns 
seen in Fig. \ref{fig:sp_200_400} are quite common in our entire 3414 pulse sequence, there 
are regions of disturbed drifting, as well as significant sections of nulls. An illustrative 
300 pulse sequence of this kind is shown in Fig. \ref{fig:sp_600_900}. Changes in drift rate,
including curved drift bands (e.g. pulse \# 720 to pulse \# 780) are seen clearly. In this 
context this pulsar behaves similar to B0826$-$34. In Fig. \ref{fig:sp_600_900} some instances 
of the pulsar in null state are also seen $-$ pulse sequences, 647 to 687, 693 to 705 and 780 
to 835. 

We note that the inner and the outer drift regions are quite closely spaced with almost no 
discontinuity between the drift bands of the two region. Furthermore, the two drift regions
are clearly locked to each other in phase $-$ the subpulse emission from the inner drift 
region is in phase with that from the outer drift region on the right side, and at the same 
time the emission in inner drift region is out of phase with the outer drift region situated 
on the left side. This phase locked relationship (hereafter PLR) is maintained for the entire 
stretch of the data and does not appear to get perturbed after intermittent nulling or during 
changes in the drift rate. 

We can characterise the observed drift pattern by estimating $P_{3}^{m}$ (the measured time 
interval between the recurrence of successive drift bands at a given pulse longitude) and 
$P_{2}^{m}$ (the measured longitude separation between two adjacent drift bands). In practice, 
the $P_{3}^{m}$ and $P_{2}^{m}$ values need not correspond to the true values  ($P_{3}^{t}$ and 
$P_{2}^{t}$) $-$ see \cite{Gupta_etal}.

\begin{figure} 
\begin{center}
\includegraphics[angle=-90, width=0.38\textwidth]{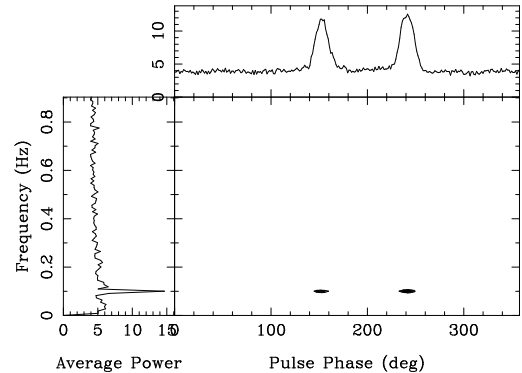}
\caption{The contour plot of the power spectrum of the flux as a function of pulse phase
during a sequence of 200 pulses (pulse \# 200$-$400). The left panel shows the power 
spectrum integrated over the entire pulse longitude. The upper panel shows the power 
integrated over fluctuation frequency.}
\label{fig:flusp_full}
\end{center}
\end{figure}       
\begin{figure}     
\begin{center}
\includegraphics[angle=-90, width=0.38\textwidth]{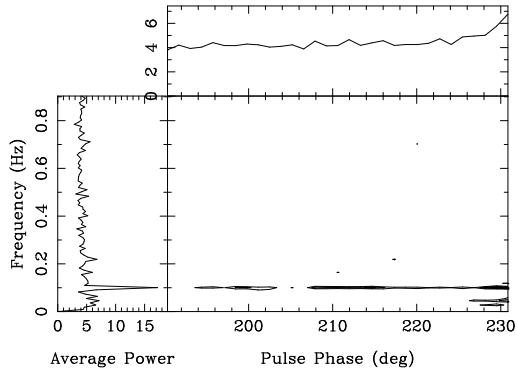}
\caption{Same as Fig. 4, but for the inner drift region only}
\label{fig:flusp_inner}
\end{center}
\end{figure}     
\begin{figure}    
\begin{center}
\includegraphics[angle=-90, width=0.38\textwidth]{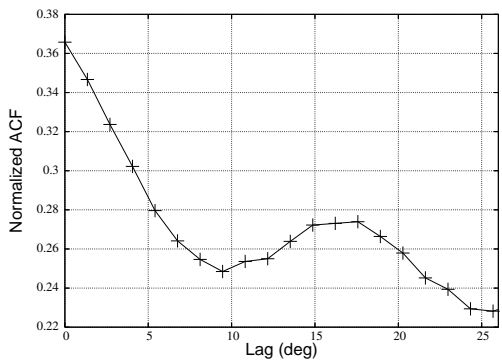}
\caption{Auto correlation results for the inner region. Secondary maxima of the 
auto correlation function near the lag of 17.5 degrees.}
\label{fig:ccf}
\end{center}
\end{figure}       

To determine $P_3^m$, we use the fluctuation spectrum analysis technique (\cite{Backer}). 
Fig. \ref{fig:flusp_full} shows the phase resolved fluctuation spectrum for the 200 pulse 
sequence of Fig. \ref{fig:sp_200_400}. There is one strong peak at $18.3\pm1.6\:P_1$ 
(where $P_1$ is the period of the pulsar). This feature is identified as $P_3^{m}$ and the 
error bar in $P_3^m$ is proportional to the reciprocal of the total length of the pulse 
sequence. Fig. \ref{fig:flusp_inner} shows the fluctuation spectrum for the inner drift 
region for the same pulse sequence. The $P_{3}^{m}$ is found to be the same for both the 
inner and the outer drift regions, which is expected given the PLR. 

The $P_{2}^{m}$ value for the inner drift region is calculated from the auto correlation 
function of the single pulses, averaged over the total number of pulses. The result is 
shown in Fig. \ref{fig:ccf}, where the secondary peak is due to the correlation between 
the adjacent drift bands and the corresponding $P_{2}^{m}$ value is 17.5\degr$\pm$1.3\degr. 
Because of the fact that one is not seeing simultaneous multiple subpulses from the outer 
drift region, $P_{2}^{m}$ for this region can not be estimated using the correlation 
analysis method. Rough estimation of $P_{2}^{m}$ value for the outer region $-$ obtained 
from the slope of the drift bands in this region $-$ is around 28\degr. Thus although the 
$P_3^m$ values are the same, the $P_2^m$ values are significantly different for the 
two drift regions.
  
\section{ Interpretation \& Modelling}               \label {sec:Interpretations }    
We interpret our observations and analysis results within the frame work of 
\cite {Ruderman_etal} model and improved versions of the same (e.g. \cite {Gil_etal_03} 
and the references therein). According to these, subpulse drifting is produced from a 
system of sub-beams (subpulse associated plasma columns). Sparks (sparking discharges 
within the vacuum gap), rotating around the magnetic axis under the action of an E x B 
drift, gives rise to a circulating pattern of sub-beams, and the time for one full 
circulation is referred to as the carousel rotation period, which we designate as $P_4$. 
As pulsar radiation beams are widely believed to be arranged in concentric cones, it it 
natural to expect the circulating sparks to be distributed in annular rings on the polar 
cap (e.g. \cite {Gil_etal_00}), each ring giving rise to one cone in the nested cones of 
emission. To confirm and support our interpretations, we have carried out simulations of 
the expected radiation pattern for this pulsar. These are described at the end of this 
section, but their results are alluded to at the different stages of our interpretation, 
which are as follows.  

(1) The possibility that the observed drift pattern could be produced from sparks 
circulating in one ring can be ruled out on the basis of the large difference in 
$P_2^m$ values and the pattern of intensity distribution between the drift regions. 
A simple interpretation of the observed drift pattern is that the inner saddle 
region corresponds to an inner conal ring that is somewhat tangentially grazed by 
the observer's line of sight (LOS), while the outer drift regions are produced by 
the intersection of the LOS with an outer conal ring. The simulation results show 
that the inner drift region can be produced by the intersection of the LOS with 
three/four neighbouring sparks of the inner ring, whereas the outer drift region 
can be created by the intersection of the LOS with one spark of the outer ring, on 
each side of the inner ring.

\begin{figure} 
\begin{center}
\includegraphics[angle=0, width=0.44\textwidth]{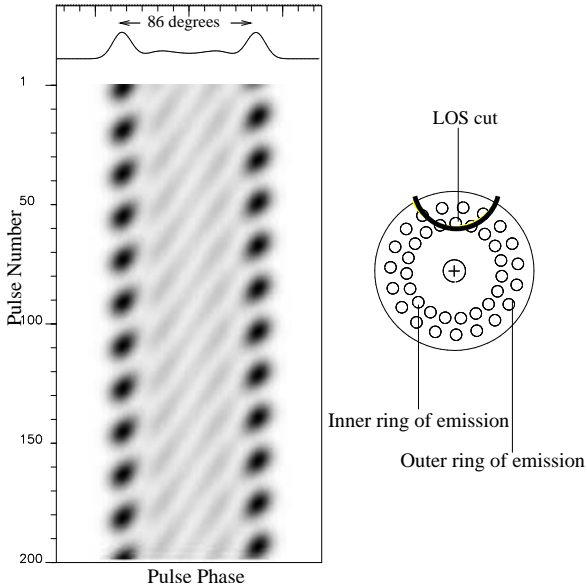}
\caption{Simulation of the subpulse drift pattern with simple dipolar geometry 
for the case: $\alpha=-11$\degr, $\beta=-5.4$\degr; drift rate$=20\degr/P_1$.}
\label{fig:sim}
\end{center}
\end{figure}

(2) The closely spaced inner and outer drift regions can be explained with a rapid 
transition of the LOS from one ring to the other. For the inner LOS geometry 
(negative $\beta$), the LOS traverses the region between the inner and the outer 
region almost perpendicularly and as a consequence one can achieve quite closely 
spaced drift regions with reasonably well separated rings of emission. This is 
borne out by the simulation results (see Fig. \ref{fig:sim}).

(3) The observed PLR implies that the apparent angular drift rate is the same for 
both the inner and the outer rings. This can be achieved with two possibilities: 
different parts of polar cap plasma rotating at the same rate (i.e.  the rotation is 
quasi-rigid), or different parts of the polar cap plasma circulating with different 
speeds which are fine tuned to maintain the apparent PLR. We do not see a natural 
way to achieve the later and hence, we suggest that the carousel rotation period, 
$P_4$, and the number of sparks, $N_{sp}$, are the same for both the inner and the 
outer ring. The simulations with two rings of emission with same $P_4$ and $N_{sp}$, 
with 180\degr out of phase emission between the rings, successfully reproduces the PLR. 

(4) What are the likely values for $N_{sp}$, $P_3^t$ (and hence $P_4$), for this pulsar?
To start with, we note that for a LOS grazing the inner ring, the $P_2^m$ is expected to 
be close to $P_2^t$. So, $N_{sp} = 360/P_2^t \approx 19-22$. Now, for $P_3^t$, there are 
two possibilities : (i) The $P_3^m$ represents an unaliased true drift rate i.e., in 
successive pulses, the observer sees the same spark, shifted to a nearby longitude.  In 
this case, $P_3^t = P_3^m = 18.3\:P_1$ and $P_4=N_{sp} \times P_3^m \approx 370\:P_1$. 
(ii) The $P_3^m$ is an aliased version of $P_3^t$ i.e., in successive pulses, a given 
spark drifts by an amount very close to the separation between adjacent sparks (or a 
multiple thereof, depending on the order of the aliasing).
As shown by \cite{Gupta_etal}, under the effect of aliasing, $P_3^t$ and $P_3^m$ are 
related as,
\begin{equation}     
\frac{1}{P_{3}^{t}} = k \frac{1}{P_{1}}+(-1)^{l} \frac{1}{P_{3}^{m}}  ~~~,
\label{eqn1}
\end{equation}  
where $k=INT[(n+1)/2]$ and $l=mod(n,2)$ depend on the alias order, $n$.  Note that we 
observe a negative drift rate for this pulsar. For an inner LOS geometry, this can happen 
for the case of an unaliased drift (as in case (i) above) $-$ the ExB drift of the spark 
plasma is slower than the rotation of the star (\cite{Ruderman}).  For an aliased drift,
this can happen when $n=2,4,6\ldots$ (i.e $k=1,2,3\ldots; l=0)$.  Taking the case for the 
lowest possible alias order ($n=2, k=1, l=0$), and using $P_3^m = 18.3\:P_1$, we obtain 
$P_3^t = 18.3/19.3 \:P_1 = 0.95\:P_1$ and
$P_4 = N_{sp}\:P_3^t = (18-21)\: P_1 \approx P_3^m$. This leads to the interesting result that 
the carousel rotation rate is the same as the measured $P_3^m$. For this scenario, the entire 
spark pattern rotates with an angular velocity of $360\degr/P_4 \approx 20\degr/P_1$.

To confirm these interpretations, we simulate the radiation from B0818$-$41 using the 
method described in \cite{Gil_etal_00}. The goal of the simulations is to reproduce the 
following observed properties : (i) width of the pulse profile, (ii) the relative 
intensities of the drift regions and (iii) the overall pattern of drifting.  Based on 
our interpretation, we use 
19 equi-spaced sparks in each of 2 concentric rings, rotating with a drift rate of 
$20\degr/P_1$. The other free parameters in the simulations are the angle between 
the magnetic and rotation axes ($\alpha$), the angle between the magnetic axis and 
the LOS ($\beta$), the radii of the rings, the sizes and relative intensities of the 
sparks. We note that the presence of closely spaced drift regions with widely different 
$P_2^m$ values, and our inference that the LOS intersects only one spark from the outer 
ring, all favour an inner LOS geometry (negative $\beta$) over an outer LOS geometry 
(positive $\beta$).

For an inner geometry, we find satisfactory results for a fairly narrow range of values 
of $\alpha$, $\beta$: roughly from $\alpha=12.2$\degr  and $\beta=-6$\degr  to $\alpha=7.5$\degr 
and $\beta=-3.35$\degr. Fig. \ref{fig:sim} shows a typical simulation result, for the case 
of $\alpha= 11$\degr and $\beta=-5.4$\degr. Here, the radius of the outer ring is 0.8 $r_p$ 
and the inner ring is 0.6 $r_p$ ($r_p$ being the polar cap radius). As can be seen, most 
of the observed features of the average profile and the drift bands are well reproduced, 
including the significant change of $P_2^m$ between the inner and the outer drift regions, 
which is clearly a geometrical effect. Interestingly, a 180\degr phase shift between 
the spark locations on the inner and outer rings is needed to match the PLR of the observed 
drift pattern (see Fig. \ref{fig:sp_200_400}).  
For an outer LOS geometry, typical best fit results are for $\alpha=2.8$\degr  and $\beta=7$\degr.
In this case, we find that for any reasonable geometry, (i) to reproduce the closely spaced 
inner and outer drift regions, the radial separation between the two rings needs to be rather 
small (0.72 and 0.82 $r_p$); (ii) to reproduce the observed $P_2^m$ in the inner region (with 
the observed profile width), the sizes of the sparks need to be very small; (iii) furthermore, 
the wide difference in $P_2^m$ values of the two regions is difficult to reproduce accurately. 

Finally, to investigate the relationship between $P_4$ and $P_3^m$, simulations have been 
tried using sparks of unequal intensities to bring out the presence of $P_4$ explicitly in 
the fluctuation spectrum. Different values of $P_4$ are achieved by using slightly different 
values of the drift rate. In the general case, two different features can be seen in the 
spectrum, corresponding to $P_3^m$ (caused by periodic drifting features) and $P_4$ (caused 
by the amplitude modulation due to the unequal sparks). For the appropriate choice of drift 
rate (equal to 20\degr/$P_1$), these two features merge/overlap in the spectrum to produce the 
measured $P_3^m$ signal. For the real data, we believe that if the $P_4$ is present in the spectrum, 
it is co-located with the aliased $P_3$ signal, at $P_3^m$. 
         
\section{Discussions and Summary}                \label {sec:Discussions} 
The remarkable subpulse drift pattern of B0818$-$41 can be interpreted as being due to 
an unaliased drift with a $P_4 \approx 370\:P_1 \approx 200 s$, or an aliased drift with 
$P_4 \approx 18.3\:P_1 \approx 10 s$. We believe that the latter is a more likely scenario 
for this pulsar. An unaliased drift can not explain the occurrence of longitude stationary 
sub-pulses and sense reversal of the drift rate that is seen in our data. For an aliased 
drift rate, this can be explained by small variations of the drift rate that move the $P_3$ 
across the nearest Nyquist boundary (\cite{Gupta_etal}). Though it is rare for the signature 
of carousel rotation period to be directly present as a low frequency feature in the 
fluctuation spectrum, we note that this has been claimed for at least two other cases : 
B0834+06 (\cite {Asgekar_etal}) and B0943$+$10 in Q mode (\cite {Asgekar_01} and \cite {Rankin_etal}).

It is interesting to compare our results with the theoretical models for $E \times B$ drift, 
such as that given by \cite{Gil_etal_00}. According to their equations (12) and (13), the 
complexity parameter $a$ (which can be represented as $Int(a) \approx 2\nu+1$, with $\nu$ being 
the putative number of cones), can be used to estimate the number of sparks in a ring as 
$N_{sp} = \pi a = P_4/P_3^t$. Taking $P_4 = m\:P_1$, we get : $P_3^t/ P_1= m\ \pi\:(2\nu+1)$. 
Further, if $P_4 = P_3^m$, then from Eqn. 1 we obtain, $P_3^t/ P_1 = m\ (k\:m\pm1)$. Hence, 
$\pi\:(2\nu+1) = (k\:m\pm1)$, which relates the number of cones to the measured periodicity 
and the alias order. Here, the plus sign applies for even alias orders, i.e. 
$n=2,4,6\cdots$, $l=0$ and the minus sign for the odd alias orders, i.e. $n=1,3,5\cdots$, $l=1$. 
Thus, for our measured periodicity of $18.3\pm1.6\:P_1$, the right hand side evaluates to 
$19.3 \pm 1.6$ for the inner LOS solution and first allowed alias order ($n=2$). For the 
outer LOS solution, it evaluates to $17.3 \pm 1.6$ for the first allowed alias order ($n=1$). 
The left hand side evaluates to $15.7 \pm 1.6$ for $\nu=2$ and to $21.9 \pm 1.6$ for $\nu=3$. 
Thus, the outer LOS solution is more compatible with the pulsar having two cones, whereas the 
inner LOS solution is in closer agreement with a three cone model. It is quite likely that the 
LOS misses the third innermost cone in this pulsar. Note that drift with higher alias orders 
(e.g. $n=3,4$) is incompatible with the above picture : either the number of sparks has to 
be twice as many (which is unlikely), or the $P_4$ has to be shorter by a factor of 2.

Furthermore, the complexity parameter can be used to estimate the screening factor $\eta$ 
for the partially screened gap model, which describe the inner acceleration region in 
pulsars (\cite {Gil_etal_03}). From equation (2) of \cite{Gil_etal}, $P_{4}/{P_1} = a / {2\eta}$. 
For unaliased drifting ($P_4 \approx 370\:P_1$), $\eta \approx 0.016$ and $\eta \approx 0.023$ 
for $\nu=2$ and $\nu=3$, respectively. For aliased drifting with $P_4 = 18.3\:P_1$, the corresponding 
estimates for $\eta$ are 0.14 and 0.19. In \cite {Gil_etal_03}, calculated $\eta$ values for different 
pulsars range from 0.032 to 0.36, with a typical value of $\approx 0.2$ meaning that on the average 
the actual potential drop is of the order of 10\% of the pure vacuum gap (\cite {Ruderman_etal}. The 
$\eta$ value for the aliased drift case for B0818$-$14 is comparable to this, whereas for the unaliased 
drift case it will be the lowest amongst the known values of $\eta$. This lends further support to our 
preference for aliased drift in the case of this pulsar.

The attempt to determine $n$ and $P_4$ has been successful only for a few other pulsars : 
B0834$+$06 ($P_4 = 14.8\:P_1 \approx 18.9\:s$, \cite {Asgekar_etal}), B0943$+$10 
($P_4 = 37\:P_1 \approx 40\:s$, $n$=0, \cite {Desh_etal}), B0809$+$74 ($P_4 = 165\:P_1 \approx 200\:s$, 
$n=0$, \cite{Leeuwen_etal}) and B0826$-$34 ($P_4 \approx 14\:P_1 \approx 25.9\:s$, $n=2$ 
\cite{Gupta_etal}).  We note that compared to these values, our results for B0818$-$41 
give $ P_4 \approx 200\:s $ for the unaliased case and $P_4 \approx 10\:s$ for the aliased case.
 
The permanent PLR between the inner and the outer drift regions is a unique feature in 
B0818$-$41. It suggests that the relative location of the circulating spark
pattern in the two rings is strongly correlated, rather than being independent. Furthermore,
the circulation rate is the same for both the rings, indicating a common electodynamic control
in the entire polar cap. Arrangement of sparks located 180\degr phase shifted in the inner and the 
outer rings would correspond to the sparks that are maximally packed on the polar cap. Spark 
discharges occur in every place where the potential drop is high enough to ignite and develop 
pair production avalanche (\cite{Ruderman_etal}), so sparks populate the polar cap as densely as 
possible. At the same time such arrangement of sparks should be quasi rigid, a property which is 
also suggested by the data. These results could be of significant implication for the physics of 
the polar cap. Multiple drift regions, with their drifts phase related to each other, are known in 
only a few pulsars, e.g. B0815$+$09 by \cite {McLaughlin_etal_04}, and B1839$-$04 by 
\cite{Weltevrede_etal}. The mirrored drift bands for B0815$+$09 are explained in the sparking gap model 
as emission coming from an inner core gap (we refer as vacuum gap) and inner annular gap 
(\cite {Qiao_etal_04}). The closely spaced inner and the outer drift regions can constrain the 
radial distances between the rings of emission in the polar cap of B0818$-$41. The radius of 
the inner and the outer ring from the simulation (Fig. \ref{fig:sim}) are, $0.6 \:r_p$ and 
$0.8 \:r_p$ $-$ are roughly in the range predicted for other pulsars by \cite {Gupta_etal_03} 
with a separate investigation method.

To summarize, we find that B0818$-$41 is a wide profile pulsar with unique drifting properties: 
closely spaced inner and outer drift regions, with apparently different drift rates, but permanently 
locked in phase to each other.  To explain the observations, we find that the pulsar needs to be a 
fairly aligned rotator, with emission seen from two conal rings on the polar cap, likely viewed with 
an inner LOS geometry. Each ring has about 19 - 21 sparks, and it is very likely that the carousel 
rotation period is about $18.3\:P_1 \approx 10\:s$ for this pulsar.
The observations and analysis presented here constitute only a preview of the remarkable 
properties of B0818$-$41.  Follow up studies, including polarisation information and single pulse 
observations at different frequencies, are expected to provide new and interesting results.\\

\noindent{\large\bf Acknowledgements :} 
We thank the staff of the GMRT for help with the observations.  The GMRT is run 
by the National Centre for Radio Astrophysics of the Tata Institute of Fundamental 
Research. We thank R. Smits for the fluctuation spectrum analysis routine, and J. Rankin
and D. Mitra for insightful discussions.  J. G. and M. S. acknowledge a partial support 
polish grant 1 P03D02926.

\label{lastpage}

\begin{thebibliography}{}
\bibitem[Asgekar \& Deshpande (2001)]{Asgekar_01} Asgekar, A., Deshpande, A. A., 2001, {\it MNRAS}, {\bf 326}, 1249.
\bibitem[Asgekar \& Deshpande (2005)]{Asgekar_etal} Asgekar, A., Deshpande, A. A., 2005, {\it MNRAS}, {\bf 357}, 1105.
\bibitem[Backer (1970)]{Backer} Backer, D. C., 1970, {\it Nature}, {\bf 227}, 692.
\bibitem[Deshpande \& Rankin (1999)] {Desh_etal} Deshpande, A. A. and Rankin, J. M., 1999, {\it ApJ}, {\bf 524}, 1008.
\bibitem[Esamdin et al. (2005)] {Esamdin_etal} Esamdin, A., Lyne, A.G., Graham-Smith, F., Kramer, M. and 
Manchester, R.N., 2005, {\it MNRAS}, {\bf 356}, 59.
\bibitem[Gil \& Sendyk (2000)] {Gil_etal_00} Gil, J. and Sendyk, M., 2000, {\it ApJ}, {\bf 541}, 351.
\bibitem[Gil, Melikidze \& Geppert (2003)] {Gil_etal_03} Gil, J., Melikidze G. I., Geppert U., 2003, {\it A\&A}, 
{\bf 407}, 315.
\bibitem[Gil et al. (2006)] {Gil_etal} Gil, J., Melikidze, G. and Zang, B., 2006, {\it A\&A}, {\bf 457}, L5.
\bibitem[Gupta et al. (2004)]{Gupta_etal} Gupta, Y., Gil, J., kijak, J. and Sendyk, M., 2004, {\it A\&A}, {\bf 426}, 229.
\bibitem[Gupta \& Gangadhara. (2003)]{Gupta_etal_03} Gupta, Y., Gangadhara, R. T., 2003, {\it ApJ}, {\bf 584}, 418.
\bibitem[Gupta et al. (2000)]{Gupta_etal_00} Gupta, Y., Gothoskar, P. B., Joshi, B. C., Vivekanand, M., Swain, R., Sirothia, S., and Bhat, N.D.R., 2000, in IAU Colloq. 177, Pulsar Astronomy, ed. M. Kramer, N. Wex, and R. Wielebinski (ASP Conf. Ser. 202; San Francisco: ASP), 277.
\bibitem[McLaughlin et al. (2004)]{McLaughlin_etal_04} McLaughlin, M. A., in IAU Symp. 218, Young Neutron Stars and Their Environments, ed F. Camilo \& B. M. Gaensler (San Francisco: ASP), 127.
\bibitem[van Leeuwen et al. (2003)]{Leeuwen_etal} van Leeuwen A. G. J., Stappers B. W., Ramchandran R., Rankin J. M., 
2003, {\it A\&A}, {\bf 399}, 223.
\bibitem[Manchester et al. (1978)]{Manchester_etal} Manchester, R. N., Lyne, A. G., Taylor, J. H., Durdin, J. M., 
Large, M.I., Little, A.G., 1978, {\it MNRAS}, {\bf 185}, 409.
\bibitem[Qiao et al. (1995)]{Qiao_etal} Qiao Guojun, Manchester, R. N., Lyne, A.G., Gould, D. M., 1995, {\it MNRAS}, 
{\bf 274}, 572.
\bibitem[Qiao et al. (2004)]{Qiao_etal_04} Qiao G., Lee, K. J., Zang, B., Xu, R. X. and Wang, H. G., 2004, {\it ApJ}, 
{\bf 616}, L127.
\bibitem[Rankin \& Suleymanova (2006)]{Rankin_etal} Rankin, J. M., Suleymanova, S. A., 2006, {\it A\&A}, {\bf 453}, 679.
\bibitem[Ruderman \& Sutherland (1975)]{Ruderman_etal} Ruderman, M. A. and Sutherland, P. G., 1975, {\it ApJ}, 
{\bf 196} 51.
\bibitem[Ruderman (1976)]{Ruderman} Ruderman, M. A., 1976, {\it ApJ}, {\bf 203} 206.
\bibitem[Weltevrede et al. (2006)]{Weltevrede_etal} Weltevrede, P., Edwards, R. T. and Stappers, B. W., 2006, 
{\it A\&A}, {\bf 445} 243.

\end{thebibliography}
\end{document}